\documentclass[twocolumn,amsmath,showpacs,amsfonts,aps,prc,floatfix]{revtex4}
\usepackage{graphicx}
\usepackage{bm}

\newcommand{\vih}{\left (\frac{v_2}{\epsilon} \right )^{ih}}
\newcommand{\etas}{\left ( \frac{1}{\tau_i T_i} \frac{\eta}{s} \right )}
\newcommand{\la}{\langle}
\newcommand{\ra}{\rangle}

\begin{document}

\title{Knudsen number, ideal hydrodynamic limit for elliptic flow and QGP viscosity in $\sqrt{s_{NN}}$=62 and 200 GeV Cu+Cu/Au+Au collisions} 
  
\author{A. K. Chaudhuri}
\email[E-mail:]{akc@veccal.ernet.in}
\affiliation{Variable Energy Cyclotron Centre, 1/AF, Bidhan Nagar, 
Kolkata 700~064, India}

\begin{abstract}
Taking into account of entropy generation during evolution of a viscous fluid,  
we have estimated inverse Knudsen number, ideal hydrodynamic limit for elliptic flow and QGP viscosity to entropy ratio in $\sqrt{s_{NN}}$=62 and 200 GeV Cu+Cu/Au+Au collisions.  Viscosity to entropy ratio is estimated as $\eta/s=0.17\pm 0.10\pm 0.20$, the first error is statistical, the second one is systematic. 
In a central Au+Au collision, inverse Knudsen number is $\approx 2.80\pm 1.63$,  which presumably small for complete equilibration. In peripheral collisions it is even less. 
Ideal hydrodynamic limit for elliptic flow is $\sim$40\% more than the experimental flow in a central collision.  
 \end{abstract}


\date{\today}  

\maketitle

Recent experiments  at RHIC produced convincing evidences that a collective matter is created  
in Au+Au collisions \cite{BRAHMSwhitepaper,PHOBOSwhitepaper,PHENIXwhitepaper,STARwhitepaper}. The evidences come mainly from observing finite elliptic flow in
non-central collisions, which is now regarded as a definitive signature of collective effect  \cite{Ollitrault:1992bk,Poskanzer:1998yz}. 
Elliptic flow  measure the azimuthal correlation of produced particles with respect to the reaction plane.  It is also best understood in a collective model like hydrodynamics \cite{QGP3}. 
In a non-central collision, the reaction zone is spatially asymmetric. Differential pressure gradient convert the spatial asymmetry in to momentum asymmetry. Ideal hydrodynamics has been quite successful explaining  a large part of the experimental elliptic flow \cite{QGP3}. However  it is now realized that the experimentally measured scaling of integrated $v_2$ with multiplicity or with collision centrality
is not in agreement with ideal hydrodynamics. While ideal hydrodynamics predicts approximate
scaling \cite{Song:2008si}, in experiments scaling is violated \cite{Voloshin:2007af,Voloshin:2006wi}. As discussed in \cite{Bhalerao:2005mm}, violation of the scaling can be understood as an indication of incomplete thermalization. Ideal hydrodynamics require local thermal equilibration.
Deviation from the local equilibrium can lead to a characteristic dependence of the eccentricity scaled elliptic flow on charged particles multiplicity. 
Degree of thermalization in the fluid produced in Au+Au collisions can be characterized by the dimensionless parameter, Knudsen number ($K$) \cite{Bhalerao:2005mm}. 
By definition, inverse of the Knudsen number is the number of collisions per particles,

\begin{equation} \label{eq2}
K^{-1}=\frac{\bar{R}}{\lambda}=\bar{R} n\sigma
\end{equation}

\noindent where $\bar{R}$ is the characteristic size of the system, $n$ is the particle density and $\sigma$ is the inter particle cross section.   Validity of hydrodynamics require that $K^{-1} >>1$, so that large number of collisions can bring the system to local thermal equilibrium. The opposite limit, $K^{-1} << 1$ is the Knudsen regime where hydrodynamics become inapplicable. The simple formula,

\begin{equation} \label{eq3}
\left ( \frac{v_2}{\epsilon} \right )^{ex}=\vih
\frac{K^{-1}}{K^{-1}+K_0^{-1}},
\end{equation}

\noindent proposed in \cite{Bhalerao:2005mm} give qualitatively correct behavior of the experimental elliptic flow. 
In the limit of small Knudsen number, experimental flow approaches the ideal hydrodynamic limit $\vih$ with a correction   linear in $K$. In the other extreme limit of large $K$, flow is proportional to the Knudsen number $K$. In Eq.\ref{eq3},
$K_0^{-1}$ is a number of the order of unity, whose precise value can be determined only from explicit transport calculations. 
From Monte-Carlo simulation of transport equations $K_0$ was estimated,  $K_0=0.70\pm 0.03$ \cite{Gombeaud:2007ub}. In \cite{Bhalerao:2005mm} it was argued that inverse of the Knudsen number $K^{-1}$ can be determined from the experimental data, as it is proportional to 
$\frac{1}{S}\frac{dN}{dy}$, where $\frac{dN}{dy}$ is the total multiplicity density and $S$ is a measure of the transverse area of the collision zone,

\begin{equation} \label{eq4}
\frac{1}{K}=c_s \sigma \frac{1}{S}\frac{dN}{dy}
\end{equation}

In Eq.\ref{eq4}, $c_s$  is the speed of sound of the medium and $\sigma$ is the inter-particle cross section. Eq.\ref{eq3} and \ref{eq4} connect two experimental observables, elliptic flow and particle multiplicity  and can be used to determine 
unknown quantities e.g. ideal hydrodynamic limit of elliptic flow $\vih$,  the combination of parameters $K_0\sigma c_s$.

However, there is a serious flaw in the derivation of Eq.\ref{eq4} and the conclusions derived from it could be misleading. Eq.\ref{eq4} 
  was obtained with the assumption that 
the total particle number is conserved throughout the evolution \cite{Bhalerao:2005mm}. The assumption is
justified in an isentropic expansion, i.e. one dimensional evolution of ideal fluid, when entropy density ($s$) times the proper time ($\tau$) is a constant. Under such condition,   $\frac{1}{S} \frac{dN}{dy} \propto s\tau \approx n \tau$ \cite{Hwa:1985xg}. However, in a viscous evolution, entropy is generated and initial and final state entropy are not same and the assumption is clearly violated.  Only in systems with very small viscosity, the assumption may be approximately valid, but not in systems where sufficient entropy is generated. 
Explicit  numerical simulations indicate that
in Au+Au collisions, entropy generation can be substantial, e.g. $\sim$ 20\%, 30\% and 50\% 
in fluid evolution with  viscosity to entropy ratio, $\eta/s$=0.08, 0.12 and 0.16  \cite{Chaudhuri:2009uk}. 
One may argue that unlike in explicit numerical simulation of viscous hydrodynamics, Knudsen number ansatz does not require the entire dynamic range of evolution.
Time of validity of Knudsen number ansatz is $\tau < {\overline R}/c_s$ \cite{Bhalerao:2005mm}. For $c_s=\sqrt{1/3}$, and characteristic size $\overline{R}\approx$ 1-2 fm, the time scale   is $\tau <$1.7-3.5  fm. 
Though the time scale is small compared to the entire dynamic range of evolution,
it is large enough for significant entropy generation.
In explicit simulation of viscous hydrodynamics, entropy generation is fast and most of the entropy is generated with first 2-4 fm of evolution \cite{Chaudhuri:2006jd} . Thus even though Knudsen number ansatz does not require full dynamic range of evolution, in the time scale for validity of Knudsen ansatz most of the entropy will be generated.

\begin{figure}[t]
\center
 \resizebox{0.30\textwidth}{!}{%
  \includegraphics{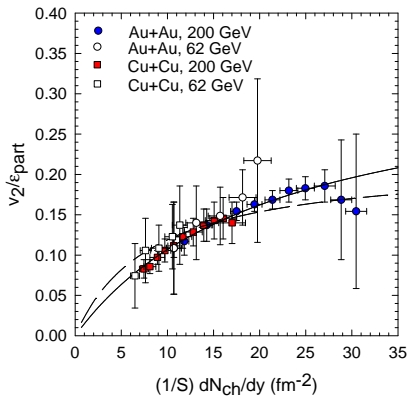}
}
\caption{(color online) PHOBOS data for the centrality dependence of eccentricity scaled elliptic flow in Cu+Cu and Au+Au  collisions at $\sqrt{s_{NN}}$= 62 and 200 GeV. The solid     line is the fit to the data with $\vih$=0.33 and $\etas$=0.83. The dashed line is the fit in the ideal fluid approximation, $\vih=0.19$ and $\etas$=0.} \label{F1}
\end{figure} 

In the present paper accounting for the entropy generation in viscous evolution, we generalise Eq.\ref{eq4}. 
 As in \cite{Bhalerao:2005mm}, we also consider one dimensional Bjorken longitudinal expansion.  If $\eta$ and $\zeta$ is the shear   and bulk viscosity coefficients, for Bjorken flow, energy-momentum conservation equation, $\partial_\mu T^{\mu\nu}=0$ reduces to   \cite{Danielewicz:1984ww,Muronga:2001zk}, 

\begin{equation} \label{eq5}
\frac{d(s \tau)}{d\tau}=\frac{1}{\tau T}( \frac{4}{3}\eta+\zeta)
\end{equation}

\noindent where $s$ is the entropy density and $\tau$ is the proper time.   In general bulk viscosity is much less than the shear viscosity, however, near the phase transition region bulk viscosity can be large  \cite{Kharzeev:2007wb,Karsch:2007jc}.  In the following we neglect the bulk viscosity. We further assume that shear viscosity is proportional to cube of the temperature. For    $\eta \propto T^3$, Eq.\ref{eq5} can be analytically integrated between time $\tau_i$ and $\tau_f$.
For $\tau_f >> \tau_i$, the final state entropy can be written as \cite{Danielewicz:1984ww,Muronga:2001zk},
 
 \begin{equation} \label{7a}
  \tau_f s_f \approx \tau_i s_i \left [1+\frac{2}{3\tau_iT_i}
  \left(\frac{\eta}{s}\right) 
  \right ]^3
\end{equation}
 
 In Eq.\ref{7a}, $T_i$ is the temperature at the time scale $\tau_i$.  
Equating final state entropy density $s_f$ with particle multiplicity per unit transverse area ($S$), $\tau_f s_f\approx 3.6\frac{1}{S}\frac{dN}{dy}$ \cite{Hwa:1985xg},  we obtain the   particle density at the time scale $\tau_i$  as,

\begin{equation} \label{eq7}
n_i \approx \frac{1}{\tau_i}\frac{1}{S} \frac{dN}{dy} \left [1+\frac{2}{3\tau_iT_i}
  \left(\frac{\eta}{s}\right)  \right ]^{-3}
\end{equation}

Inserting Eq.\ref{eq7} in Eq.\ref{eq2}, in the time scale $\tau_i \leq \bar{R}/c_s$, i.e. as long as the transverse size of the system does not vary significantly \cite{Bhalerao:2005mm},
inverse of Knudsen number can be obtained as,

\begin{equation} \label{eq8}
\frac{1}{K} \approx  
 \sigma c_s    \left [\frac{1}{S}\frac{dN}{dy} \right ]  
\left [1+\frac{2}{3\tau_iT_i} \left(\frac{\eta}{s}\right)   \right ]^{-3} 
 \end{equation}
 
One immediately observes that neglect of entropy generation during evolution will   over estimate  $K^{-1}$, by the factor $\left [1+\frac{2}{3\tau_iT_i} \left(\frac{\eta}{s}\right)   \right ]^{3}$. As it will be shown below, experimental data indicate that the factor could be large, $\sim$ 2-7.
 
Inserting Eq.\ref{eq8} in Eq.\ref{eq3}, eccentricity scaled elliptic flow now can be related to observed particle multiplicity as,  

   \begin{equation} \label{eq9}
\left (\frac{v_2}{\epsilon} \right )^{ex}=\left (\frac{v_2}{\epsilon} \right )^{ih} 
\frac{  \frac{1}{S}\frac{dN}{dy} \left [1+\frac{2}{3\tau_iT_i} \left(\frac{\eta}{s}\right)  \right ]^{-3}  }
{\frac{1}{K_0 \sigma c_s} + \frac{1}{S}\frac{dN}{dy} \left [1+ \frac{2}{3\tau_iT_i} \left(\frac{\eta}{s}\right)  \right ]^{-3}  
}
\end{equation} 

Experimental data on elliptic flow and particle multiplicity can be fitted with Eq.\ref{eq9} to obtain  estimates of  the hydrodynamic limit of elliptic flow $\vih$  and viscosity to entropy ratio in unit of initial time and temperature $\etas$.   
 Eq.\ref{eq9} also involve the quantity $K_0\sigma c_s$. $K_0$, $\sigma$ and $c_s$ are known approximately. For example, inter parton cross section is expected to be small, $\sigma$=3-4 mb. From transport calculations,  $K_0$ was estimated as $K_0=0.7\pm 0.3$. The speed of sound of QGP medium is expected to be $c_s\approx\sqrt{1/3}$. In the following, we fix $\sigma$=3 mb, $K_0$=0.7 and $c_s=\sqrt{1/3}$.

\begin{figure}[t]
\center
 \resizebox{0.30\textwidth}{!}{%
  \includegraphics{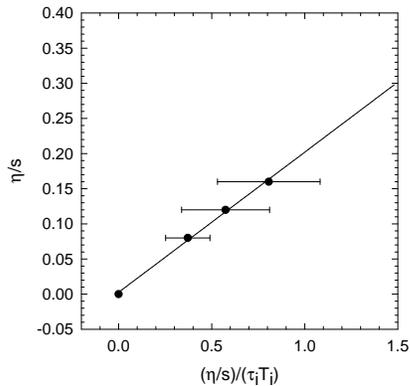}
}
\caption{Viscosity to entropy ratio $\eta/s$ as a function of the parameter $\etas$ 
from the analysis in   \cite{Chaudhuri:2009uk}.
The solid line is a straight line fit.}
 \label{F2}
\end{figure}   



In Fig.\ref{F1},   PHOBOS \cite{Back:2004mh,Alver:2006wh,Alver:2008zza} data for the centrality dependence of (participant) eccentricity scaled charged particles elliptic flow, in Au+Au and Cu+Cu collisions at $\sqrt{s_{NN}}$=62 and 200 GeV  are shown.  
PHOBOS data  have large error bars and within the error data do not show any system size or energy dependence. All the four data sets are fitted together with Eq.\ref{eq9}. The solid line in Fig.\ref{F1} shows the fit. Data are well explained, $\chi^2$ value is also small, $\chi^2/N\approx 0.1$. However, $\vih$ and $\etas$ can be determined only with large uncertainties,  $\vih=0.33\pm 0.12$, $\etas=0.83\pm 0.51$. Uncertainty in $\vih$ and $\etas$ would be reduced with better quality data. While definitive conclusions cannot be made due to large uncertainty in $\vih$,     the central value suggests that   even in central Au+Au collisions, ideal hydrodynamic limit is not reached. In central/mid-central collisions, experimental flow  
is only $\sim$ 60\% of the ideal fluid limit. In peripheral collisions, it is even less.  
For comparison purpose, in Fig.\ref{F1}, we have shown the fit to the data   in the ideal fluid approximation, $\etas=0$.  In the ideal fluid approximation ideal hydrodynamic limit for elliptic flow is estimated as, $\vih= 0.19\pm 0.04$. However, description to the data is comparatively poor, $\chi^2/\approx 1.3$  is $\sim$ 10 times larger than that obtained in viscous evolution. 
 
\begin{table}[t] 
\caption{\label{table1} Inverse Knudsen number ($K^{-1}$), in ideal and viscous fluid,   as a function of collision centrality in Au+Au collision. Also listed are charged particles multiplicity per unit transverse area ($\frac{1}{S}\frac{dN_{ch}}{dy}$) from PHOBOS experiment \cite{Back:2004mh,Alver:2006wh,Alver:2008zza} and characteristic size $\bar{R}$ of the system.  Inter parton cross section is assumed to be $\sigma$= 3 mb, speed of sound $c_s=\sqrt{1/3}$.
} 
\begin{ruledtabular} 
  \begin{tabular}{|c|c|c|c|c|}\hline
 &   &  & \multicolumn{2}{c|}{$K^{-1}$}   \\  \hline
collision  &   $\frac{1}{S}\frac{dN_{ch}}{dy}$ & $\bar{R}$ &$\etas=0.83\pm0.51$ & $\etas$=0 \\  
centrality(\%)  & $(fm^{-2})$ & (fm) &   &\\ \hline  
0-3&        30.46 & 2.01& $2.80\pm1.63$& 7.91    \\
3-6&        28.84 & 1.93& $2.65\pm1.55$& 7.49     \\
6-10&       27.06 & 1.84& $2.49\pm1.45$& 7.03     \\
10-15&      24.95 & 1.75& $2.29\pm1.33$& 6.48    \\
15-20&      23.17 & 1.64& $2.13\pm1.24$& 6.02     \\
20-25&      21.39 & 1.55& $1.97\pm1.15$& 5.56    \\
25-30&      19.44 & 1.45& $1.79\pm1.04$& 5.05   \\
30-35&      17.50 & 1.38& $1.61\pm0.94$& 4.55   \\
35-40&      14.26 & 1.30& $1.31\pm0.76$& 3.70    \\
40-45&      11.83 & 1.23& $1.09\pm0.63$& 3.07    \\
45-50&      10.69 & 1.18& $0.98\pm0.57$& 2.78     \\
\end{tabular}\end{ruledtabular}  
\end{table}  

With the estimate of $\etas$, we can compute inverse Knudsen number from Eq.\ref{eq8}. In table.\ref{table1}, we have listed $K^{-1}$ for different centrality ranges of Au+Au collisions. Particle multiplicities are taken from the PHOBOS data \cite{Back:2004mh,Alver:2006wh,Alver:2008zza}.  For comparison, we have also listed the values in the ideal fluid approximation. $K^{-1}$ decreases by a factor of 2-7 when entropy generation is accounted for.
In the ideal fluid approximation, $K^{-1}\approx 8$ in a central collision.
It is reduced to $\sim$2.8 if entropy generation during evolution is accounted for. While it is debatable whether $K^{-1}\approx 8$ can lead to complete equilibration, it is unlikely that complete thermalization
will be achieved with $K^{-1}\approx 2.8$.   In more peripheral collisions, equilibration is certainly incomplete.
 
\begin{table}[t]
\caption{\label{table2} Listed are some estimates of QGP viscosity to entropy ratio from experimental data in Au+Au collisions at RHIC. The observables analyzed are also listed.}
  \begin{ruledtabular} 
  \begin{tabular}{|c|c|c|}\hline
Sl. no.  &   $4\pi \eta/s$ & Experimental observable  \\ 
         \hline\hline
1 & $0.88 \pm 0.38 \pm 1.76$  & $\phi$ meson's $\la N\ra$, $\la p_T\ra$ and $v_2$ \cite{Chaudhuri:2009uk}\\
2 & 1.0-3.77 & $p_T$ fluctuations \cite{Gavin:2006xd}\\         
3 & 1.4-2.4 & $v_2$ scaling violation \cite{Drescher:2007cd} \\  
4 & $\approx$2 & $v_2$ scaling violation \cite{Masui:2009pw} \\  
5 & $1.13\pm 0.19 \pm 1.26$ & $v_2$ scaling violation \cite{Lacey:2006bc}\\
6 & $1.51\pm 0.38$ & $v_2$ scaling violation \cite{Chaudhuri:2009ud}\\
7 & 1.3-2.0 & heavy quark energy loss \cite{Adare:2006nq}\\
8 & $1.45\pm 0.06$ & $v_2$ \cite{Chaudhuri:2009hj}\\
9 & $\leq$ 1.51 & $p_T$ spectra of $\pi$, $K$ and $p$ \cite{Roy:2010qg}\\
10 & $2.14\pm 1.26\pm 2.56$\footnote{present work} &   $v_2 \& \frac{1}{S}\frac{dN_{ch}}{dy} $  \\
\end{tabular}\end{ruledtabular}  
\end{table}   

QGP viscosity is an important parameter characterizing QGP medium.  String theory based models (ADS/CFT) give a lower bound on viscosity to entropy ratio of any matter, $4\pi \eta/s \geq  1$ \cite{Policastro:2001yc}. In \cite{Luzum:2008cw,Song:2008hj}, from experimental data, a phenomenological upper bound    was conjectured, $4\pi \eta/s < 5$. We have obtained viscosity to entropy ratio in unit of initial time and temperature. It can be converted to more comprehensible viscosity to entropy ratio if the initial time and temperature scale  is known. Recently, 
in \cite{Chaudhuri:2009uk} STAR data on $\phi$ mesons multiplicity, mean $p_T$ and integrated $v_2$ were analyzed in ideal and viscous fluid dynamics. At the initial time $\tau_i$=0.6 fm,   ideal or viscous fluid was initialized to reproduce experimental $\phi$ meson multiplicity. Viscous fluid requires less initial temperature than an ideal fluid. Results of the analysis are shown in Fig.\ref{F2}, where $\eta/s$ as a function of the parameter  $\etas$ is shown.  The solid line in Fig.\ref{F2} is a straight line fit, $\eta/s =0.2\etas$. The relation can be used to  convert extracted $\etas$  to $\eta/s$. We obtain, $\eta/s=0.17\pm 0.10$. From the PHOBOS data, viscosity to entropy ratio can be determined only within $\sim$ 60\% accuracy. Evidently, much better quality  data are required for
more precise determination of viscosity to entropy ratio. As mentioned earlier, the estimate was obtained with $K_0=0.7$, $c_s=\sqrt{1/3}$ and $\sigma=0.3 fm^2$ corresponding to, $K_0\sigma c_s\approx 0.121$. Estimate of viscosity depend on the value of $K_0\sigma c_s$.   Systematic uncertainty in $\eta/s$ is $\sim$ 120\% due to a factor of 2 uncertainty in  $K_0\sigma c_s$. We then estimate 
QGP viscosity to entropy ratio as $\eta/s=0.17\pm 0.10 \pm 0.20$, the first error is statistical, the second is systematic. Systematic uncertainty will increase if uncertainty in initial time and temperature scale is included. The estimated value is well with the two bounds, $1 \leq 4\pi \eta/s \leq 5$ \cite{Policastro:2001yc,Luzum:2008cw,Song:2008hj}.
In table.\ref{table2}, present estimate for QGP viscoisty is compared with some recent estimates. 
One may note that presently estimated  $\eta/s$ is similar to the values obtained in previous extractions \cite{Drescher:2007cd},\cite{Masui:2009pw}, which disregarded  entropy generation in the Knudsen ansatz. The reason is understood. Experimental data include the effect of entropy generation. Thus even if viscous effects are  neglected in the Knudsen ansatz, 
the fitted  paratmeters $\vih$ and $K_0\sigma c_s$ (see Eq.\ref{eq9}) will include the effect. 

We note that the present estimate should be considered as an upper limit for QGP viscosity. We have  neglected bulk viscosity. Experimental data   include the effect of bulk viscosity. Neglect of bulk viscosity will be compensated by  increasing $\eta/s$. Also, we have neglected transverse expansion. Experimental data also include the effect of transverse expansion. One observes from Eq.\ref{eq8} that   $K^{-1}$ will decrease if transverse expansion is included (transverse area of system at freeze-out will be larger than the initial area). Neglect of the transverse expansion will be compensated again by  increasing $\eta/s$.

To conclude, taking into account that entropy is generated during evolution of a viscous fluid, we have generalized a relation between inverse Knudsen number $K^{-1}$  and    particle density $\frac{1}{S}\frac{dN}{dy}$. PHOBOS data on the centrality dependence of elliptic flow indicate that  $K^{-1}$ is overestimated    by a factor $\sim$ 2-7 if entropy generation is neglected. We have also estimated  ideal hydrodynamic limit for elliptic flow $\vih$ and QGP viscosity
to entropy ratio $\eta/s$. Estimated $\vih$ is $\sim$40\% larger than the scaled flow in a central collision. Estimated viscosity to entropy ratio, $4\pi \eta/s=2.14\pm 1.26$ is well within
the ADS/CFT lower bound and phenomenological upper bound, $1 \leq 4\pi \eta/s \leq 5$ .

\end{document}